\documentclass[12pt]{article}
\usepackage{setspace, amsmath, graphicx, multirow, bigdelim, float, amsfonts, caption, color, colortbl, lscape, subcaption,bm, array,natbib,url}

\usepackage{amssymb, amsthm}
\usepackage{bm}
\usepackage{subcaption}
\usepackage{setspace}
\usepackage{array, booktabs}

\begin{document}
\title{Data Fusion for Correcting Measurement Errors}

\author{Tracy Schifeling, Jerome P. Reiter, Maria DeYoreo\thanks{This research was supported by The National Science Foundation under 
award SES-11-31897. The authors wish to thank Seth Sanders for his input on informative prior specifications, and Mauricio Sadinle for discussion that improved the strategy for accounting for the informative sample design.}}
\date{}
\maketitle

\begin{abstract}

\noindent
Often in surveys, key items are subject to measurement errors. Given just the data, it can be difficult 
to determine the distribution of this error process, and hence to obtain accurate inferences that involve the error-prone variables.  In some settings, 
however, analysts have access to a data source on different individuals with high quality measurements of the  
error-prone survey items.  We present a data fusion framework for leveraging this information to improve inferences in the error-prone survey.
The basic idea is to posit models about the rates at which individuals make errors, coupled with 
models for the values reported when errors are made. This can avoid the unrealistic assumption of
 conditional independence typically used in data fusion. We apply the approach  
on the reported values of educational attainments in the American Community Survey, using  
the National Survey of College Graduates as the high quality data source.  In doing so, we account for the informative sampling design used to select the National Survey of College Graduates.  We also present a
process for assessing the sensitivity of  various analyses to different choices for the measurement error models.  Supplemental material is available online.
\end{abstract}


\section{Introduction}
Survey data often contain items that are subject to measurement errors. For example, some respondents might misunderstand a question or accidentally select the wrong response, thereby providing values unequal to their factual values.  Left uncorrected, these measurement errors can result in degraded inferences \citep{kim2015}. Unfortunately, the distribution of the measurement errors typically is not estimable from the 
survey data alone.  One either needs to make strong assumptions about the measurement error process \citep[e.g., as in ][]{curran2009}, or leverage information from some other source of data, as we do here.  

One natural source of information is a validation sample, i.e., a dataset with both the reported, possibly erroneous values and the true values measured on the same individuals.  These individuals could be a subset of the original survey
\citep{pepe1992, yucel2005},
or a completely distinct set \citep{raghu2006, schenker2007, schenker2010, Carrig:2015}. With validation data, one can model the relationship between the error-prone and true values, and use the model to replace the error-prone items with multiply-imputed,  plausible true values \citep{Reiter:Biometrika, siddique}.

In many settings, however, it is not possible to obtain validation samples, e.g., because it is too expensive or because someone other than the analyst collected the data.  In such cases, another potential source of information is a separate, ``gold standard'' dataset that includes true (or at least very high quality) measurements of the items subject to error, but not the error-prone measurements.  Unlike validation samples, the gold standard dataset alone does not provide enough information to estimate the relationship between the error-prone and true values; it only provides information about the distribution of the true values. Thus, analysts are faced with a special case of data fusion \citep{rubinstatmatch, moriarity, rassler, dorazio,  reiter:datafusion,fosdick2015}, i.e., integrating information from two databases with disjoint sets of individuals and distinct variables.

One default approach, common in other data fusion contexts, is to assume that the error-prone and true values are conditionally independent given some set of variables $\bm{X}$ common to both the survey and gold standard data.  Effectively, this involves using the gold standard data to estimate a predictive model for the true values from $\bm{X}$, and applying the estimated model to impute replacements for all values of the error-prone items in the survey.  However, this conditional independence assumption completely disregards the information in the error-prone values, which sacrifices potentially useful information.  For example, consider national surveys that ask people to report their educational attainment.  We might expect most people to report values accurately and only a modest fraction to make errors. It does not make sense to alter every individual's reported values in the survey, as would be done using a conditional independence approach.

In this article, we develop a framework for leveraging information from gold standard data to improve inferences in surveys subject to measurement errors.  The basic idea is to encode plausible assumptions about the error process, e.g., most people do not make errors when reporting educational attainments, and the reporting process, e.g., when people make errors, they are more likely to report higher attainments than actual, into statistical models.  We couple those models with distributions for the underlying true data values, and use multiple imputation to create plausible corrections to the error-prone survey values, which then can be analyzed using the methods from \citet{rubin:1987}.  This allows us to avoid unrealistic conditional independence assumptions in lieu of more scientifically defensible models.

The remainder of this article is organized as follows. In Section~\ref{sec:litreview}, we review an example of misreporting of educational attainment in data collected by the Census Bureau, so as to motivate the methodological developments. 
In Section~\ref{sec:model}, we introduce the general framework for specifying measurement error models to leverage the information in gold standard data.  
 In Section~\ref{sec:app}, we apply the framework to handle potential measurement error in educational attainment in the 2010 American Community Survey (ACS), using the 2010 National Survey of College Graduates (NSCG) as a gold standard file.  
In doing so, we deal with a key complication in the data integration: accounting for the informative sampling design used to sample the NSCG. 
We also demonstrate how the framework facilitates analysis of the sensitivity of conclusions to different measurement error model specifications.  In Section~\ref{sec:discussion}, we provide a brief summary.

\section{Misreporting in Educational Attainment}\label{sec:litreview}

To illustrate the potential for reporting errors in educational attainment that can arise in surveys,
we examine data from the 1993 NSCG. The 1993 NSCG surveyed individuals who indicated 
on the 1990 census long form that they had at least a college degree \citep{ncses12201}.
The questionnaire asked about educational attainment, including detailed 
 questions about educational histories.  These questions greatly reduce the possibility of respondent error, so that the educational attainment values in the NSCG can be considered a gold standard \citep{sanders2003}.
The census long form, in contrast, did not include detailed follow up questions, so that reported educational attainment is prone to measurement error.

The Census Bureau linked each individual in the NSCG to their corresponding record in the long form data.
The linked file is available for download from the Inter-university Consortium for Political and Social Research \citep{icpsrdata}.
Because of the linkages, we can characterize the actual measurement error mechanism for educational attainment in the 1990 long form data.  In the NSCG, we  
treat the highest degree of the three most recent degrees reported (coded as ``ed6c1'', ``ed6c2'', and ``ed6c3'' in the file) as the true education level. 
 We disregard any degrees earned in the years 1990--1993, as these  
occur in the three year gap between collection of the long form and NSCG data.  This ensures consistent time frames for the NSCG and long form reported values. 
We cross-tabulate these degrees with the degrees reported in the long form data (coded ``yearsch'' in the file).
Table~\ref{table:NSCG} displays the cross-tabulation.
A similar analysis was done by \citet{sanders2003}.

\begin{table}[t]
\caption[Unweighted cross-tabulation of NSCG-reported and Census-reported education from the 1993 NSCG linked dataset.]{Unweighted cross-tabulation of reported education in the NSCG and census long form from the linked dataset.  BA stands for bachelor's degree; MA stands for master's degree; Prof stands for professional degree; and, PhD stands for Ph.\ D.\ degree.  The 14,319 individuals in the group labeled No Degree did not have a college degree, despite reporting otherwise.  The 51,396 individuals 
in the group labeled Other did not have one of (BA, MA, Prof, PhD) and are discarded from subsequent analyses.}\label{table:NSCG}
\begin{center}
\begin{tabular}{l r | r | r | r | r | r r}
& \multicolumn{1}{r}{}& \multicolumn{4}{c}{Census-reported education} &\\
& \multicolumn{1}{r}{}& \multicolumn{4}{c}{$\overbrace{\hspace{5cm}}$} &\\
& \multicolumn{1}{r}{}& \multicolumn{1}{l}{BA} & \multicolumn{1}{l}{MA} & \multicolumn{1}{l}{Prof} & \multicolumn{1}{l}{PhD} & \textit{Total}\\ \cline{3-6}
\ldelim\{{4}{1cm}[\parbox{2cm}{NSCG-reported education}] & BA & 89580 & 4109 & 1241 & 249 & 95179 \\ \cline{3-6}
& MA & 1218 & 33928 & 655 & 526 & 36327\\ \cline{3-6}
& Prof & 382 & 359 & 8648 & 563 & 9952\\ \cline{3-6}
& PhD & 99 & 193 & 452 & 6726 &  7470\\ \cline{3-6}
& \multicolumn{1}{r}{\textit{Total}} &\multicolumn{1}{r}{91279} & \multicolumn{1}{r}{38589}& \multicolumn{1}{r}{10996}& \multicolumn{1}{r}{8064}& \multicolumn{1}{r}{148928}\\
& \multicolumn{6}{r}{} \\ \cline{3-6}
& No Degree  &10150 & 1792& 2040& 337 & 14319\\ \cline{3-6}
& Other & 33368 & 10912& 4710& 2406 & 51396\\ \cline{3-6}
\end{tabular}
\end{center}
\end{table}

As evident in Table~\ref{table:NSCG}, reported education levels on the long form often are higher
 than those on the NSCG, particularly for individuals with only a bachelor's degree. 
Of the 163,247 individuals in scope in the NSCG,  over 14,000 were determined not to have at least a bachelor's 
 degree when asked in the NSCG, despite reporting otherwise in the long form.
A whopping 33\% of individuals who reported being professionals in the long form actually 
are not professionals according to the NSCG. One possible explanation for this error is confusion over the definition of professionals.  The Census 
Bureau intended the category to capture graduate degrees from universities (e.g., J.D., M.B.A, M.D.), whereas \citet{sanders2003} found that individuals in professions such as cosmetology, nursing, and health services, which require certifications but not graduate degrees, selected the category.

In spite of the nontrivial reporting error, the overwhelming majority of individuals' reported education levels are consistent in the long form and in the NSCG. Of the individuals in the NSCG who had at least a college degree at the time of the 1990 census, 
about 93.3$\%$ of them have the same contemporaneous education levels in both files. This suggests that most people
report correctly, an observation we want to leverage when constructing measurement error models for education in the 2010 ACS.

In most situations, we do not have the good fortune of observing individuals' error-prone and true values simultaneously.  Instead, we are in the setting represented by Figure~\ref{fig:unlinked}.  This is also the case in our analysis of educational attainments in the 2010 ACS, described in Section~\ref{sec:app}.  The sampling frame for the 2010 NSCG is constructed from  
reported education levels in the ACS, which replaced the long form (after the 2000 census). 
However, unlike in 1993, linked data are not available as public use files.   Therefore, we  treat the 2010 NSCG as gold standard data, 
and posit measurement models that connect the information from the two data sources, using the framework that we now describe.

\begin{figure}[t]
\begin{center}
\begin{tabular}{ r | c | c | c |}
 \multicolumn{1}{c}{}& \multicolumn{1}{c}{$\bm{X}$} &  \multicolumn{1}{c}{$Y$} & \multicolumn{1}{c}{$Z$} \\ \cline{2-4}
 \ldelim\{{4}{.85cm}[$D_E$] & \multirow{4}{*}{$\checkmark$} &  \multirow{4}{*}{?} & \multirow{4}{*}{$\checkmark$} \\
 & & & \\
& & & \\
 & & & \\ \cline{2-4}
\ldelim\{{2}{.85cm}[$D_G$] & \multirow{2}{*}{$\checkmark$} &  \multirow{2}{*}{$\checkmark$} & \multirow{2}{*}{?} \\
 & & & \\ \cline{2-4}
\end{tabular}

\end{center}
\caption[Graphical representation of data fusion set-up.]{
Graphical representation of data fusion set-up. In the survey data $D_E$, we only observe the error-prone measurement $Z$ but not the true value $Y$. In the gold standard data $D_G$, we only observe $Y$ but not $Z$. We observe variables $\bm{X}$ in both samples.} \label{fig:unlinked}
\end{figure}

\section{Measurement Error Modeling via Data Fusion}\label{sec:model}

As in Figure~\ref{fig:unlinked}, let $D_E$ and $D_G$ be two data sources comprising distinct individuals, with sample sizes  
$n_E$ and $n_G$, respectively. For each individual $i$ in $D_G$ or $D_E$, let $\bm{X}_i = (X_{i1}, \ldots, X_{ip})$ be variables common to both surveys, such as demographic variables. We assume these variables have been harmonized \citep{dorazio}  across $D_G$ and $D_E$ and are free of errors.  Let $Y$ represent the error-free values of some variable of interest, and let
$Z$ be an error-prone version of $Y$. We observe $Z$ but not $Y$ for the $n_E$ individuals in $D_E$.  We observe $Y$ but not $Z$ for the $n_G$ individuals in $D_G$.   For simplicity of notation, we assume no missing values in any variable, although the multiple imputation framework easily handles missing values. Additionally, $D_E$ can include variables for which there is no corresponding variable in $D_G$.  These variables do not play a role in the measurement error modeling, although they can be used in multiple imputation inferences.

We seek to estimate $\mathrm{Pr}(Y,Z \mid \bm{X})$, and use it to create multiple imputations for the missing values in $Y$ for the individuals in $D_E$.  We do so for the common setting where $(\bm{X}, Y, Z)$ are all categorical variables; similar ideas apply for other data types. For $j=1, \dots, p$, let each $X_j$ have $d_j$ levels.
Let $Z$ have $d_Z$ levels and $Y$ have $d_Y$ levels. Typically $d_Z = d_Y$, but this need not be the case generally.  For example, in the NSCG/ACS application, $Z$ is the educational attainment among those who report a college degree in the ACS, which has $d_Z = 4$ levels (bachelor's degree, master's degree, professional degree, or Ph.\ D.\ degree), and 
$Y$ is the educational attainment in the NSCG, which has $d_Y = 5$ levels. An additional level is needed because some individuals in the NSCG truly do not have a college degree.

For all $i \in D_E$, let $E_i$ be an (unobserved) indicator of a reporting error, that is, $E_i = 1$ when $Y_i \neq Z_i$ and $E_i = 0$ otherwise. Using $E$ enables us to write $\mathrm{Pr}(Y,Z \mid \bm{X})$ as a product of three sub-models. For individual $i$, the full data likelihood (omitting parameters for simplicity) can be factored as  
\begin{multline}
\mathrm{Pr}(Y_i = k, Z_i = l \mid \bm{X}_{i}) = \mathrm{Pr}(Y_i = k \mid \bm{X}_i) \\
\times \mathrm{Pr}(E_i = e | Y_i = k, \bm{X}_i) \mathrm{Pr}(Z_i = l | E_i = e,  Y_i = k, \bm{X}_i).\label{jointmodel}
\end{multline}
This separates the true data generation process and the measurement error generation process, which facilitates model specification. In particular, we can use $D_G$ to estimate the true data distribution $\mathrm{Pr}(Y\mid \bm{X})$.  We then can posit 
different models for the rates of making errors, $\mathrm{Pr}(E_i = e | Y_i = k, \bm{X}_i)$, and for the reported values when errors are made, $\mathrm{Pr}(Z_i = l | E_i = 1,  Y_i = k, \bm{X}_i)$. Intuitively, the error model locates the records for which $Y_i\neq Z_i$, and the reporting model captures the patterns of misreported $Z_i$. Of course, when $E_i=0$, $\mathrm{Pr}(Z_i = Y_i) = 1$. 
A similar factorization is  used by \citet{yucel2005},  \citet{he2014}, \citet{kim2015}, and \citet{daniel2015}, among others.

By construction, $D_G$ and $D_E$ cannot be used to estimate any of the conditional probabilities $\mathrm{Pr}(Y \mid Z, \bm{X})$ directly. Hence,  we have to restrict the number and types of parameters in the sub-models in \eqref{jointmodel}. Put another way, if we tried to estimate a fully saturated model for $(E, Z \mid \bm{X})$, we would not be able to identify all the parameters by using $D_G$ and $D_E$ alone.  To see this, assume for the moment that all $d_X = \Pi_{j=1}^p d_j$ possible combinations of $\bm{X}$ are present in $D_G$ and $D_E$. To estimate the distribution of $(E, Z \mid \bm{X})$ using a fully saturated model, we require $(d_Y-1)d_X + (d_Z -1)d_Yd_X = (d_Yd_Z-1)d_X$ independent pieces of information from $(D_G, D_E)$, where each subtraction of one derives from the requirement that probabilities sum to one.  However, $D_G$ and $D_E$ together provide only $(d_Z-1) d_X + (d_Y-1) d_X + d_X = (d_Z + d_Y - 1)d_X$ independent pieces of information, where we add a $d_X$ to properly account for the sum to one constraint. 
A key insight here is that since the true data model requires $d_Yd_X$ parameters to estimate the joint distribution for $(Y,\bm{X})$, the data can identify at most $(d_Z-1)d_X$ parameters in the error and reporting models, combined. 
Related identification issues arise in the context of refreshment sampling to adjust for nonignorable attrition in longitudinal studies \citep{hirano2001combining, refresh, si:reiter:politicalanalysis}.

\subsection{True data model $\mathrm{Pr}(Y_i = k \mid \bm{X}_i)$}
One can use any model for $(Y | \bm{X})$ that adequately describes the conditional distribution, such as a (multinomial) logistic regression. In the NSCG/ACS application,, we use a fully saturated multinomial model, accounting for the informative sampling design in $D_G$ using the approach described in Section~\ref{sec:complex}. One also could use a joint distribution for $(Y,\bm{X})$, such as a log-linear model or a mixture of multinomials model \citep{dunson:xing:2009, si:reiter:jebs}.

\subsection{Error model $\mathrm{Pr}(E_i = 1 | Y_i, \bm{X}_i)$}\label{sec:error}

In cases where $d_Y = d_Z$, a generic form for the error model is 
\begin{equation}
\mathrm{Pr}(E_i = 1 | \bm{X}_i, Y_i = k) = g(\bm{X}_i, Y_i, \beta), \label{eq:error}
\end{equation}
where $g(\bm{X}_i, Y_i, \beta)$ is some function of its arguments and $\beta$ is some set of unknown parameters.  A convenient class of functions that we use here is the logistic regression of $E_i$ on some design vector $\bm{M}_i$ derived from $(\bm{X}_i, Y_i)$, with corresponding coefficients $\beta$.  
The analyst can encode different versions of $\bm{M}_i$ to represent  assumptions about the error process.

The simplest specification is to set each $\bm{M}_i$ equal to a vector of ones, which implies that there is a common probability of  error for all individuals. 
This error model makes sense when the analyst believes the errors in $Z$ 
occur completely at random; for example, when errors arise simply because respondents accidentally and randomly select the wrong response in the survey, or when all respondents are equally likely to misunderstand the survey question.
A more realistic possibility is to allow the probability of error to depend on some variables in $\bm{X}_i$ but not on $Y_i$, e.g., men misreport education at different rates than women.  This could be encoded  by including an intercept for one of the sexes in $\bm{M}_i$. Finally, one can allow the probability of error to depend on $Y_i$ itself---for example, people who truly do not have at least a college degree are more likely to misreport---by including some function of it in $\bm{M}_i$.

In the case where $d_Z \neq d_Y$, as in the NSCG/ACS application, we automatically set $E_i=1$ for any individual with $Y_i \notin \{1:d_Z\}$. For example, we set $E_i=1$ for all individuals who are determined in the NSCG not to have a college degree but report so in the ACS.  The stochastic part of the error model only applies to individuals who truly have at least a bachelor's degree.

\subsection{Reporting model $\mathrm{Pr}(Z_i | E_i = 1,  Y_i, \bm{X}_i)$}\label{sec:reporting}
When there is no reporting error for individual $i$, i.e., $E_i = 0$, we know that $Z_i = Y_i$. When there is a reporting error, we must model the reported value $Z_i$. As with \eqref{eq:error}, one can posit a variety of distributions for the reporting error, which is some function $h(\bm{X}_i, Y_i, \alpha)$ with parameters $\alpha$.  We now describe a few reporting error models for illustration.  One could use more complicated models, e.g., based on multinomial logistic regression, as well.

A simple model  assumes that values of $Z_i$ are equally likely, as in \citet{daniel2015}. We have
\begin{equation}
\mathrm{Pr}(Z_i = l | \bm{X}_i, Y_i=k, E_i = 1) = 
\begin{cases}
1/(d_Z-1) & \text{if } l\neq k, k\in\{1:d_z\}\\
1/d_Z & \text{if } k \notin \{1:d_Z\}\\
0 & \text{otherwise}.
\end{cases} \label{eq:uniform}
\end{equation}
Such a reporting model could be reasonable when reporting errors are due to clerical errors. We note that this model does not accurately characterize the reporting errors in the 1993 linked NSCG data, per Table~\ref{table:NSCG}.

Alternatively, one can allow the probabilities to depend on $Y_i$, so that 
\begin{equation}
(Z_i | \bm{X}_i, Y_i = k, E_i = 1) \sim 
    \text{Categorical}(p_{k}(1), \ldots, p_{k}(d_Z)),
\label{eq:reportprobs}
\end{equation}
where each $p_{k}(l)$ is the probability of reporting $Z=l$ given that $Y=k$, and $p_k(k) = 0$. 
One can further parameterize the reporting model so that the reporting probabilities vary with $\bm{X}$. 
For example, to make the probabilities vary with sex and true education values, we can use 
\begin{equation}
{\small
(Z_i | \bm{X}_{i}, Y_i = k, E_i = 1) \sim
\begin{cases}
    \text{Categorical}(p_{M,k}(1), \ldots, p_{M,k}(d_Z)) & \text{if } X_{i,sex} = \text{M}\\
    \text{Categorical}(p_{F,k}(1), \ldots, p_{F,k}(d_Z)) & \text{if } X_{i,sex} =\text{F}.
\end{cases} \label{eq:reportprobsgender}
}
\end{equation}

\subsection{Specifying and estimating the model}

As apparent in Sections \ref{sec:error} and \ref{sec:reporting}, the error and reporting models can take on many 
specifications. Without linked data, analysts cannot use exploratory data analysis to inform the model choice.
Instead, we recommend that analysts posit scientifically defensible measurement error models, and make post-hoc checks 
of the sensibility of analyses from those models. We demonstrate this approach in Section \ref{sec:app}.
For example, analysts can check whether or not the predicted probabilities of errors implied by the model seem plausible.
As another diagnostic, analysts can compare the distribution of the imputed values of $(Y \mid \bm{X})$ in $D_E$ to the 
empirical distribution of $(Y \mid \bm{X})$ in $D_G$. This is akin to diagnostics in multiple imputation for missing data 
that compare imputed and observed values \citep{abayomi}. When these distributions differ substantially, it suggests the measurement error model specification (or possibly the true data model) is inadequate. 
Such diagnostic checks only can reveal problems with the model specification; they do not indicate that a particular 
specification is correct. 

More generally, it is prudent to keep the restrictions on the number of identifiable parameters in mind when specifying 
the models.  At most one can identify the equivalent of $(d_Z-1)d_X$ parameters in the combined model for $(E_i, Z_i \mid \bm{X}_i)$. 
Generally, for ease of specification and interpretation, we favor rich error models, e.g., with $\bm{M}_i$ including variables in $\bm{X}_i$ and $Y_i$, 
coupled with simple reporting models like those in Section \ref{sec:reporting}.

The exact strategy for estimating the model depends on the features of $D_G$ and $D_E$.  When both datasets
can be treated as simple random samples, we suggest using a fully Bayesian approach after concatenating $D_G$ and $D_E$.
Here, one can use typical prior distributions for the true data and error models.  For 
reporting models like those in \eqref{eq:reportprobs} and \eqref{eq:reportprobsgender}, 
it is convenient to use independent Dirichlet priors for each 
$(p_k(1), \ldots,p_k(k-1), p_k(k+1), \ldots, p_k(d_Z))$. 
In the NSCG/ACS application, we create prior distributions for the reporting models using the information from 
Table \ref{table:NSCG}.  Absent such information, analysts can use uniform prior distributions. 

When it does not make sense to concatenate $D_G$ and $D_E$, it can be convenient to use a 
multi-stage estimation strategy.  When imputing missing $Y$ in $D_E$, all of the information needed from $D_G$ is represented by the parameters of the true data model, $\theta$. Hence, we first can construct a (possibly approximate) posterior distribution of $\theta$ using only $D_G$.  We then 
sample many draws from this distribution.  We plug these draws in the Gibbs sampling steps for a Bayesian predictive 
distribution for $(Y_i \mid Z_i, \bm{X}_i, \theta)$ for the cases in $D_E$, thereby generating the multiple imputations. We describe the
Gibbs sampler for this step for the NSCG/ACS application in the supplementary material.

\section{Adjusting for Reporting Errors in Education in the 2010 ACS}\label{sec:app}

We now use the framework to adjust inferences for potential reporting error in educational attainment in the 2010 ACS, using the public use microdata for the 2010 NSCG as the gold standard file $D_G$. We consider two main analyses that could be affected by reporting error in education. First, we estimate from the ACS the number of science and engineering degrees awarded to women.  We base the estimate on an indicator in the ACS for whether or not each individual has such a degree.
Second, we examine average incomes across degrees.
This focus is motivated in part by the findings of \citet{sanders2006, sanders2008}, who found that apparent wage gaps in the 1990 census long form data could be explained by reporting errors in education.

As $D_E$, we use the subset of ACS microdata that includes only individuals who reported a bachelor's degree or higher and are under age 76.  The resulting sample size is $n_E=600,150$.  In $\bm{X}$, we include gender, age group (24 and younger, 25 - 39, 40 - 54, and 55 and older), and an indicator for whether the individual's race is black or something else. 
In the NSCG, we discarded 38 records with race suppressed, leaving a sample size of $n_G =  77,150$. 

We consider two sets of measurement error model specifications. The first set uses specifications like those in Section \ref{sec:model}, with flat prior distributions for all parameters. We use this set to illustrate model diagnostics and sensitivity analysis absent  prior information about the measurement error process.  The second set uses a common error and reporting model with different, informative prior distributions on its parameters.  We construct these informative prior distributions 
based on the analysis of the 1993 linked file.  For all specifications considered, we create $M=50$ multiple imputations of the plausible true education values in the 2010 ACS, which we then analyze using the methods of \citet{rubin:1987}. 
For all specifications, the true data model is a saturated multinomial distribution for the five values of $Y$ for each combination of $\bm{X}$. We begin by describing how we estimate the parameters of the true data distribution, accounting for the informative sampling design of the NSCG.

\subsection{Accounting for informative sampling design of NSCG}\label{sec:complex}

The 2010 NSCG uses reported education in the 2010 ACS as a stratification variable \citep{ncses12201, nscgdoc}. Its unweighted percentages can over-represent or under-represent  
degree types in the population; this is most obviously the case for individuals without a college degree ($Y_i=5$). We need to account for this informative sampling when estimating parameters of the true data model. 
We do so with a two stage approach. First, we 
use survey-weighted inferences to estimate population totals of $(Y \mid \bm{X})$ from the 2010 NSCG.  Second, we turn these estimates into an approximate Bayesian posterior distribution for input to fitting the measurement error models used to impute plausible values of $Y_i$ for individuals in the ACS. We now describe this process, which can be used generally when $D_G$ is collected via a complex survey design.

Suppose for the moment that $d_Y = d_Z$.  This is not the case when $D_E$ is the ACS (where $d_Z = 4$) and $D_G$ is the NSCG (where $d_Y = 5)$; however, we start here to fix ideas.  
For all possible combinations $\bm{x}$, let $\theta_{xk} = \text{Pr}(Y = k | \bm{X}=\bm{x})$, and let 
$\theta_{x} = (\theta_{x1},\ldots,\theta_{xd_Y})$. We seek to use $D_G$ to specify $f(\theta | \bm{X}, Y)$. 
To do so,  we first parameterize $\theta_{xk} = T_{xk}/\sum_{j=1}^{d_Y} T_{xj}$, where $T_{xk}$ is the population 
count of individuals with $(\bm{X}_i=\bm{x}, Y_i = k)$. 
We estimate $T_x = (T_{x1}, \dots, T_{xd_Y})$ and the associated covariance matrix of the estimator using standard survey-weighted estimation. Let $w_i$ be the sample weight for 
all $i \in D_G$. We compute the estimated total and associated variance for each $\bm{x}$ and $k$ as
\begin{align}
\hat{T}_{xk} &= \sum_{i=1}^{n_G} w_i I(\bm{X}_i = \bm{x}, Y_i = k) \label{eq:that} \\ 
\widehat{\text{Var}}(\hat{T}_{xk}) &= \frac{n_G}{n_G-1}\sum_{i=1}^{n_G}\left(w_i I(\bm{X}_i = \bm{x}, Y_i = k) - \frac{\hat{T}_{xk}}{n_G}\right)^2. \label{eq:thatvar}
\end{align}
For each $k$ and $l$, with $l \neq k$, we also compute the estimated covariance,\begin{multline}
\widehat{\text{Cov}}(\hat{T}_{xk}, \hat{T}_{xl}) = \frac{n_G}{n_G-1}\sum_{i=1}^{n_G}\Bigg[\left(w_i I(\bm{X}_i = \bm{x}, Y_i = k) - \frac{\hat{T}_{xk}}{n_G}\right)\\
\times \left(w_i I(\bm{X}_i = \bm{x}, Y_i = l) - \frac{\hat{T}_{xl}}{n_G}\right)\Bigg]. \label{eq:thatcov}
\end{multline}
The variance and covariance estimators are the design-based estimators for probability proportional 
to size sampling with replacement, as is typical of multi-stage complex surveys \citep{lohr}.

Switching now to a Bayesian modeling perspective, we assume that $T_x \sim$ Log-Normal($\mu_x, \tau_x$), so as to ensure a distribution with positive values for all true totals.
We select $(\mu_x, \tau_x)$ so that each E$(T_{xk}) = \hat{T}_{xk}$ and Var$(T_x) = \hat{\Sigma}(\hat{T}_x)$, the estimated covariance matrix with elements defined by \eqref{eq:thatvar} and \eqref{eq:thatcov}. 
These are derived from moment matching \citep{lognorm}. We have 
\begin{align}
\mu_{xj} &= \log(\hat{T}_{xj})-\tau_{x}\lbrack j,j\rbrack/2 \label{eq:mu}\\ 
\tau_{x}\lbrack j,j\rbrack &= \log\left(1 + \hat{\Sigma}_{x}\lbrack j,j\rbrack/(\hat{T}_{xj}^2)\right) \label{eq:tau}\\
\tau_{x}\lbrack j,i\rbrack &= \log\left(1 + \hat{\Sigma}_x\lbrack j,i\rbrack/(\hat{T}_{xj}\cdot\hat{T}_{xi})\right),\label{eq:cov}
\end{align}   
where the notation $[j,i]$ denotes an element in row $j$ and column $i$ of the matrix. We draw $T^*_x$ from this log-normal distribution, and transform to draws
$\theta^*_x$.

Since the 2010 NSCG does not include individuals who claim in the ACS to have less than a bachelor's degree, 
we cannot use $D_G$ directly to estimate ${T}_{x5}$. Instead, 
we estimate $T_{x+} = T_{x1} + T_{x2} + T_{x3} + T_{x4} + T_{x5}$ using the ACS data, and estimate $(T_{x1}, T_{x2}, T_{x3}, T_{x4})$ from 
the NSCG using the method described previously; this leads to an estimate for $T_{x5}$. More precisely, let the ACS design-based estimator for $T_{x+}$ be 
 $\hat{T}_{x+}$, with design-based variance estimate $\hat{\sigma}^2(\hat{T}_{x+})$.
We sample a value $T^*_{x+} \sim$ Normal$(\hat{T}_{x+}, \hat{\sigma}^2(\hat{T}_{x+}))$.  Using an independent sample of values of 
$(T^*_{x1}, \dots, T^*_{x4})$ from the NSCG,
we compute $T^*_{x5} = T^*_{x+} - \sum_{j=1}^4T^*_{xj}$, and set $T^*_x = (T^*_{x1}, \dots, T^*_{x5})$.   We repeat these steps 10,000 times.  We then compute the mean and covariance matrix of the 10,000 draws, which we 
again plug into \eqref{eq:mu} -- \eqref{eq:cov}.  The resulting log-normal distribution is the approximate posterior distribution of $\theta_x$. We include an example of this entire procedure in the supplementary material.

\subsection{Measurement error models}

\begin{table}[t]
\caption{Summary of the first four measurement error model specifications for 2010 NSCG/ACS analysis.  These models use flat prior distributions on all parameters.}\label{table:appmodels}
\begin{center}
\footnotesize
\begin{tabular}{r  c c }
\toprule
& Error model & Reporting model \\
& Expression for $M_i^T\beta$ &$Pr(Z_i | Y_i = k, E_i = 1)$ \\
\midrule
Model 1 &$\beta_1 + \sum_{k =2}^{4}\beta_kI(Y_i=k)$ &  $\text{Categorical}(p_{k}(1), \ldots, p_{k}(4))$ \\
\\
Model 2 & $\beta_1 + \sum_{k =2}^{4}\beta_k^{(M)}I(Y_i=k, X_{i,sex} = \text{M})$ &$\text{Categorical}(p_{k}(1), \ldots, p_{k}(4))$ \\
\\
Model 3 & 
$\beta_1 + \sum_{k =2}^{4}\beta_k^{(no)}I(Y_i=k, X_{i,black} = \text{no})$ & $\text{Categorical}(p_{k}(1), \ldots, p_{k}(4))$ \\
& $+  \sum_{k =1}^{4}\beta_k^{(yes)}I(Y_i=k, X_{i,black} = \text{yes})$ & \\
\\
Model 4 & $\beta_1 + \sum_{k =2}^{4}\beta_k^{(M)}I(Y_i=k, X_{i,sex} = \text{M})$ & $\text{Categorical}(p_{M,k}(1), \ldots, p_{M,k}(4))$ if $X_{i,sex} =$ M \\
& $+  \sum_{k =1}^{4}\beta_k^{(F)}I(Y_i=k, X_{i,sex} = \text{F})$ &  $\text{Categorical}(p_{F,k}(1), \ldots, p_{F,k}(4))$ if $X_{i,sex} =$F \\
\\
\bottomrule
\end{tabular}
\end{center}
\end{table}

The two sets of measurement error models include four that use flat prior distributions and three that use informative prior distributions based on the 1993 linked data.  For all error models, we use a logistic regression of $E_i$ on various main effects and interactions of $Y_i$ and $\bm{X}_i$. For all reporting models, we use categorical distributions with probabilities that depend on $Y_i$ and possibly $\bm{X}_i$. 
The four models with flat prior distributions are summarized in Table~\ref{table:appmodels}.
In Model 1, the error and reporting models depend only on $Y_i$.  
Model 2 and 3 keep the reporting model as in \eqref{eq:reportprobs} but expand the error model. In Model 2, the probability of a reporting error can vary with $Y_i$ and sex ($X_{i,sex}$). In Model 3, error probabilities can vary with $Y_i$ and the indicator for black race ($X_{i,black}$). In Model 4, the error and reporting models both depend on $Y$ and sex.

\begin{table}[t]
\caption{Summary of informative prior specifications for 2010 NSCG/ACS analysis for males with bachelor's degrees.}\label{table:apppriors}
\begin{center}
\footnotesize
\begin{tabular}{c  l l }
\toprule
 & Error rate  & Reporting probabilities $(p_{M,1}(2), p_{M,1}(3), p_{M,1}(4))$\\ \hline
Model 4 & Beta(1, 1) & Dirichlet(1, 1, 1)\\
Model 5 & Beta(.76, 14.24) & Dirichlet(3.54, 1.27, 0.19) \\
Model 6 & Beta(2724.2, 50862) & Dirichlet(2235.3, 799.7, 123.1)\\
Model 7 & Beta(500, 99500) & Dirichlet(1, 1, 1)\\ \bottomrule
\end{tabular}
\end{center}
\end{table}

For Models 5 -- 7,  we use the specification in Model 4 and incorporate prior information about the measurement errors from the 1993 linked data. In constructing the priors, we first remove records  that have been flagged as having missing education that has been imputed, because these imputations might not closely reflect the actual education values \citep{sanders2003}. Table \ref{table:apppriors} displays the prior distributions for males with bachelor's degrees.  Details on how we arrive at these and other groups' prior specifications are in the supplementary material; here, we summarize  briefly. 

For Model 5, we set the prior distributions for each $\beta_k^{(x)}$ so that the error rates are centered at the estimate from the 1993 linked data. We also require the central $95\%$ probability interval of the prior distribution on each error rate to be close to $(.005, .20)$, allowing for a wide but not unrealistic range of possible error rates.  For the reporting probabilities  
$p_{M,k}(z)$ and $p_{F,k}(z)$, we center most of the prior distributions at the corresponding estimates from the 1993 linked data.  We  require the central $95\%$ probability interval of each prior distribution to have support on values of $p_{\cdot,k}(z)$ within $\pm .10$ of the 1993 point estimate, truncating at zero or one as needed. One exception is the reporting probabilities for those with ``no college degree'' who report ``professional'' degree, which we center at half the 1993 estimate.  The Census Bureau has improved the clarity of the definition of Professional in the 20 years since the 1990 long form, as discussed in the prior specification section of the supplementary material. 

For Model 6, we use the same prior means as in Model 5 for both error and reporting models.  However, we substantially tighten the prior distributions to make the prior variance accord with the uncertainty in the point estimates from the 1993 linked data.  We do so by using prior sample sizes that match those from the 1993 NSCG.  For example, the 1993 NSCG included 53,586 males with bachelor's degrees (excluding those records who had their Census education imputed). We therefore use $\text{Beta}(2724.2, 50862)$ as the prior distribution for the error rate for this $\bm{x}$. We similarly increase the prior sample sizes for the reporting probabilities to match the 1993 NSCG sample sizes.

Model 7 departs from the 1993 linked data estimates and encodes a strong prior belief that almost no one misreports their education except for haphazard mistakes. Here, we set the prior mean for the probability of misreporting education to .005 for all demographic groups. We use a prior sample size of 100,000, making the prior distribution concentrate strongly around .005.  For the reporting probabilities, we use a non-informative prior distribution for convenience, since the estimates of the reporting probabilities are strongly influenced by the concentrated prior distributions on the error rates. 

Finally, for comparison purposes, we also fit the model based on a conditional independence assumption (CIA). To impute $Y_i$ for individuals in the ACS under the CIA, we sample $\theta^*$ and then impute $(Y^* | \theta^*, \bm{X})$ from the true data model.  Here, we do not use the reported value of $Z_i$ in the imputations. 

\subsection{Empirical results}
We first examine what each model suggests about the extent and nature of the measurement errors in the 2010 ACS.  We then use the models to assess sensitivity of results about the substantive questions related to number of degrees and income.

\subsubsection{Distributions of errors in reported ACS education values}

\begin{table}[t]
\caption[Results from different reporting error models applied to 2010 data.]{Error rate estimates from different model specifications. Models 1-7 are run for 100,000 MCMC iterations. We save $M=50$ completed datasets under each model. For each dataset, we compute the estimated overall error rate, estimated error rate by gender and imputed $Y$, and associated variances using ratio estimators that incorporate the ACS final survey weights. }\label{table:app}
\footnotesize
\begin{center}
\begin{tabular}{r  c  c  c  c  c }
\toprule
&\multicolumn{4}{c}{Estimate by group}&\parbox{2cm}{Estimate overall} \\ \midrule
&$Y$=BA&$Y$=MA&$Y$=Prof.&$Y$=PhD&\\
\textbf{CIA model} & \\ 
Male & .37 (.36, .37) & .76 (.75, .76) & .91 (.91, .92) & .94 (.93, .95) & \multirow{2}{*}{.57 (.55, .58)}\\
Female & .35 (.35, .36) & .72 (.71, .72) & .95 (.94, .95) & .97 (.96, .97) &\\ 
\\
\textbf{Model 1}\\
Male & .05 (.04, .06) & .10 (.08, .11) & .18 (.15, .21) & .27 (.23, .31) & \multirow{2}{*}{.17 (.16, .19)}\\
Female & .05 (.05, .06) & .09 (.08, .10) &.18 (.15, .21) & .28 (.24, .32) &\\ 
\\
\textbf{Model 2} & \\ 
Male & .05 (.04, .06) & .18 (.16, .21) & .27 (.18, .37) & .36 (.30, .42) & \multirow{2}{*}{.20 (.18, .21) }\\
Female & .05 (.05, .06) & .12 (.10, .14) & .26 (.20, .33) & .41 (.29, .53) &\\ 
\\
\textbf{Model 3} & \\ 
Male & .05 (.04, .06) &.09 (.08, .11) & .17 (.14, .20) & .25 (.21, .30) & \multirow{2}{*}{.17 (.16, .19)}\\
Female & .05 (.05, .06) & .09 (.08, .10) & .17 (.14, .20) & .26 (.21, .31) &\\ 
\\
\textbf{Model 4} & \\
Male & .05 (.04, .06) & .19 (.16, .23) & .36 (.26, .46) & .36 (.27, .45) & \multirow{2}{*}{.22 (.20, .24)}\\
Female & .09 (.08, .10) & .14 (.11, .17) & .52 (.44, .59) & .55 (.40, .70)&\\ 
\\
\textbf{Model 5} & \\  
Male & .07 (.06, .08) & .19 (.16, .22) & .23 (.14, .32) & .34 (.27, .41) & \multirow{2}{*}{.22 (.20, .24)}\\
Female & .09 (.08, .10) & .12 (.09, .15) & .50 (.43, .57) & .31 (.17, .46) & \\ 
\\
\textbf{Model 6} & \\ 
Male & .05 (.05, .05)& .09 (.08, .10) & .10 (.09, .11) & .10 (.09, .11) &\multirow{2}{*}{.16 (.14, .17)}\\
Female & .05 (.04, .05) & .06 (.05, .07) & .16 (.14, .18) & .07 (.06, .09) \\
\\
\textbf{Model 7} & \\ 
Male & .01 (.01, .01) & .01 (.00, .01) & .00 (.00, .01) & .01 (.00, .01) & \multirow{2}{*}{.11 (.09, .13)}\\
Female & .01 (.01, .01) & .01 (.01, .01) & .01 (.00, .01) & .01 (.00, .01) \\
\bottomrule
\end{tabular}
\end{center}
\end{table}

Table~\ref{table:app} displays the multiple imputation point estimates and  95$\%$ confidence intervals for
the proportions of errors by gender and NSCG education, obtained from the $M=50$ draws of $E_i$ for all individuals in $D_E$. 
We begin by comparing results for the set of models with flat prior distributions (Models 1 -- 4) and the CIA model, then move to the set of models with informative prior distributions (Models 5 -- 7).  

The CIA model suggests extremely large error percentages, especially for the highest education levels.  These rates seem unlikely to be reality, leading us to reject the CIA model. The overall error rates for Models 1 --  4 are similar and more realistic than those from the CIA model.  The differences in error estimates between Model 2 and Model 1 suggest that the probability of error depends on sex.  Comparing results for Model 3 and Model 1, however, we see little evidence of important race effects on the propensity to make errors.

Model 4 generalizes Model 2 by allowing the reporting probabilities to vary by sex. If these probabilities were similar across sex in reality, we would expect the two models to produce similar results. However, the estimated error rates are fairly different; for example, the estimated proportion of errors for female professionals from Model 4 is about double that from Model 2. To determine where the models differ most, we examine the estimated reporting probabilities, displayed in Table~\ref{table:appreport4}. Model 4 estimates some significant differences in reporting probabilities by gender. For example, males with bachelor's degrees who make a reporting error are estimated to report a master's degree with probability .96, whereas females with bachelor's degrees who make a reporting error are estimated to report a master's degree with probability .67 and a professional degree with probability .30. Other large differences exist for professional degree holders. Females with professional degrees who make a reporting error are most likely to report a bachelor's degree, whereas men with professional degrees who make a reporting error are most likely to report a master's degree or Ph.\ D.  We note that some of the estimates for Model 4 are based on small sample sizes, which explains the wide standard errors.

\begin{table}[t]
\caption{Estimated mean and 95$\%$ confidence interval of reporting probabilities under Model 2 and reporting probabilities by gender under Model 4.}\label{table:appreport4}
\begin{center}
\footnotesize
\begin{tabular}{l c c c c} 
\toprule
& $Z$=BA & $Z$ = MA & $Z$ = Prof. & $Z$ = PhD\\ \midrule
$Y$=BA\\ 
Model 2& - & .95 (.87, 1.00) & .04 (.00, .11) & .01 (.00, .03) \\
Model 4 - Male & - & .96 (.90, 1.00) & .02 (.00, .07) & .02 (.00, .05) \\
Model 4 - Female & - &  .67 (.58, .76) &.30 (.22, .38) & .03 (.00, .07) \\
\\
$Y$=MA \\
Model 2 & .02 (.00, .06) & - & .51 (.43, .59) & .47 (.39, .55) \\
Model 4 - Male & .04 (.00, .11) & - & .57 (.48, .66) & .39 (.31, .47) \\
Model 4 - Female & .11 (.00, .25) & - & .39 (.26, .52) & .50 (.40, .61)  \\
\\
$Y$=Prof. \\
Model 2 & .05 (.00, .16) & .69 (.54, .83) & - & .26 (.14, .38) \\
Model 4 - Male & .02 (.00, .06) & .69 (.44, .94) & - & .29 (.04, .54) \\
Model 4 - Female & .91 (.79, 1.00) & .06 (.00, .16) & - & .04 (.00, .10) \\
\\
$Y$= PhD \\
Model 2  & .01 (.00, .04) & .39 (.15, .63) & .60 (.36, .83) & -\\
Model 4 - Male & .01 (.00, .05) & .21 (.02, .39) & .78 (.60, .96) & -\\
Model 4 - Female & .10 (.00, .30) & .77 (.50, 1.00) & .13 (.00, .34) &- \\
\\
$Y$=None \\
Model 2 & .95 (.95, .96) & .03 (.03, .04) & .01 (.01, .01) & .00 (.00, .00) \\
Model 4 - Male & .97 (.96, .97) & .03 (.02, .03) & .01 (.00, .01) & .00 (.00, .00) \\
Model 4 - Female & .96 (.95, .97) & .04 (.03, .05) & .00 (.00, .00) & .00 (.00, .00)\\
\bottomrule
\end{tabular}
\end{center}
\end{table}

Turning to Models 5 -- 7, we can see the impact of the informative prior distributions by comparing results in 
Table \ref{table:app} under these models to those for Model 4.  Moving from Model 4 to Model 5, the
most noticeable differences are for women with a Ph.\ D. and men with a master's degree, for whom Model 5 suggests lower error rates. 
These groups have smaller sample sizes, so that the data do not swamp the effects of the prior distribution. 
When making the prior sample sizes very large as in Models 6 and 7, the information in the prior distribution tends to overwhelm the information in the data.  We provide more thorough investigation of the impact of the prior specifications in the supplementary material.

Of course, we cannot be certain which model most closely reflects the true measurement error mechanism.  The best we can do is perform diagnostic tests to see which models, if any, should be discounted as not adequately describing the observed data.  For each ACS imputed dataset $D_E^{(m)}$ under each model, we compute the sample proportions, $\hat{\pi}^{(m)}_{xk}$, and corresponding multiple imputation 95$\%$ confidence intervals for all $16\dot5$ unique values of $(\bm{X}, Y)$.  We determine how many of the 80 estimated population percentages of $Y\mid \bm{X}$ computed from the 2010 NSCG (using the estimated $\hat{T}_{x+}$ from the ACS to back into an estimate of $\hat{T}_{x5}$) fall within the multiple imputation 95$\%$ confidence intervals.  Models that yield low rates do not describe the data accurately.

For Model 1, 73 of 80 NSCG population share estimates are contained in the ACS multiple imputation intervals. Corresponding counts are 75 for Model 2, 71 for Model 3, and 76 for Model 4.  These results suggest that Model 1 and Model 3 may be inferior to Model 2 and Model 4.  For the models with informative prior distributions, the counts are 74 for Model 5, 67 for Model 6, and 54 for Model 7. Although the prior beliefs in models 6 and 7 seem plausible at first glance, the diagnostic suggests that they do not describe the 2010 data distributions as well as Models 4 and 5.

Considering the results as well as the diagnostic check, if we had to choose one model we would select Model 5. It seems plausible that the probability of misreporting education, as well as the reported value itself when errors are made, depend on both sex and true education level. Additionally, the prior distribution from the 1993 linked data pulls estimates in groups with little sample size to measurement error distributions that seem more plausible on face value.  However, one need not use the data fusion framework for measurement error to select a single model; rather, one can use the framework to examine sensitivity of analyses to the different specifications.

\subsubsection{Sensitivity analyses}

\begin{figure}[t]
\begin{subfigure}{.48\textwidth}
\centering
\includegraphics[clip,trim=1cm 6cm 1cm 6cm,scale=.4]{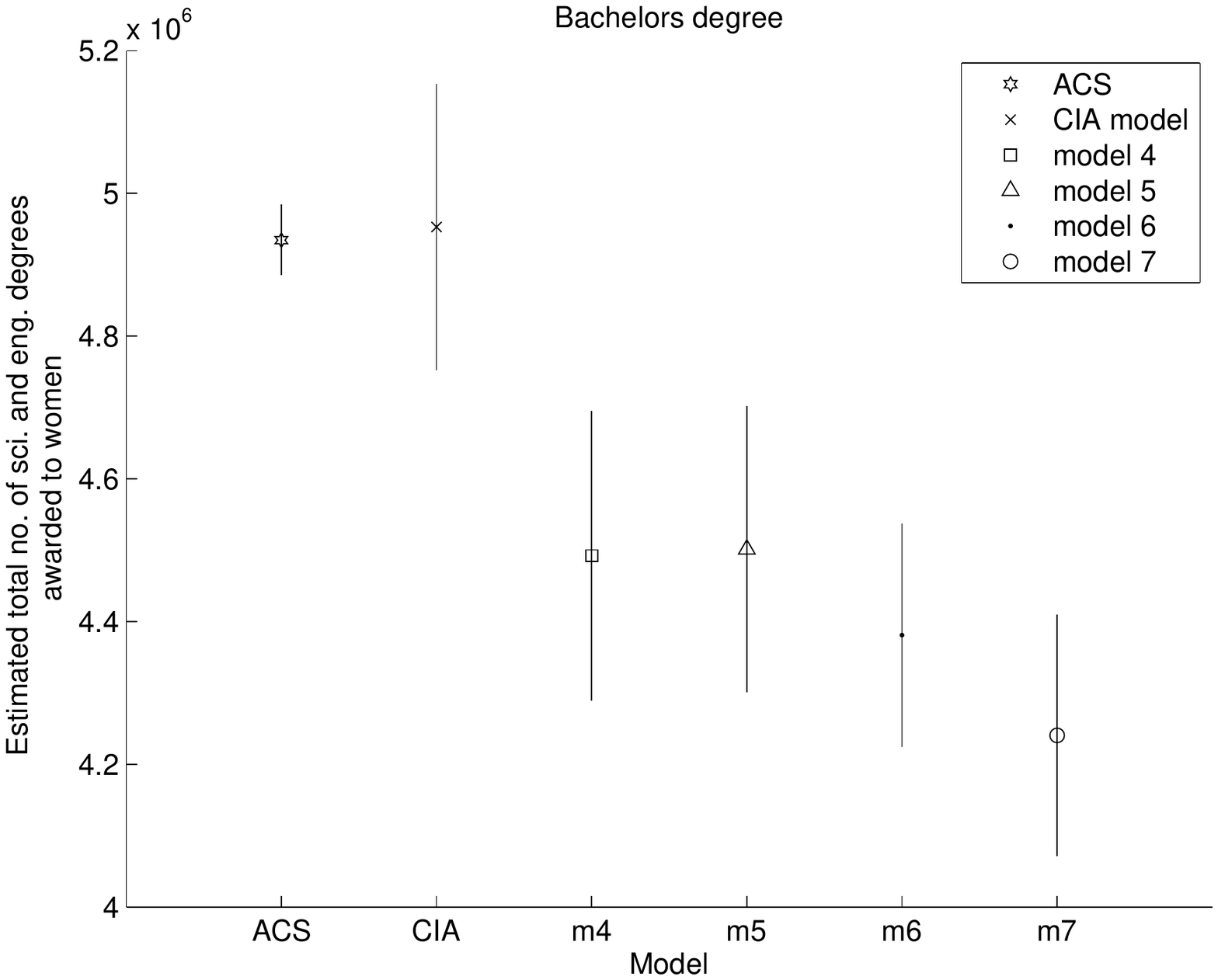}
\end{subfigure}
\begin{subfigure}{.48\textwidth}
\centering
\includegraphics[clip,trim=1cm 6cm 1cm 6cm,scale=.4]{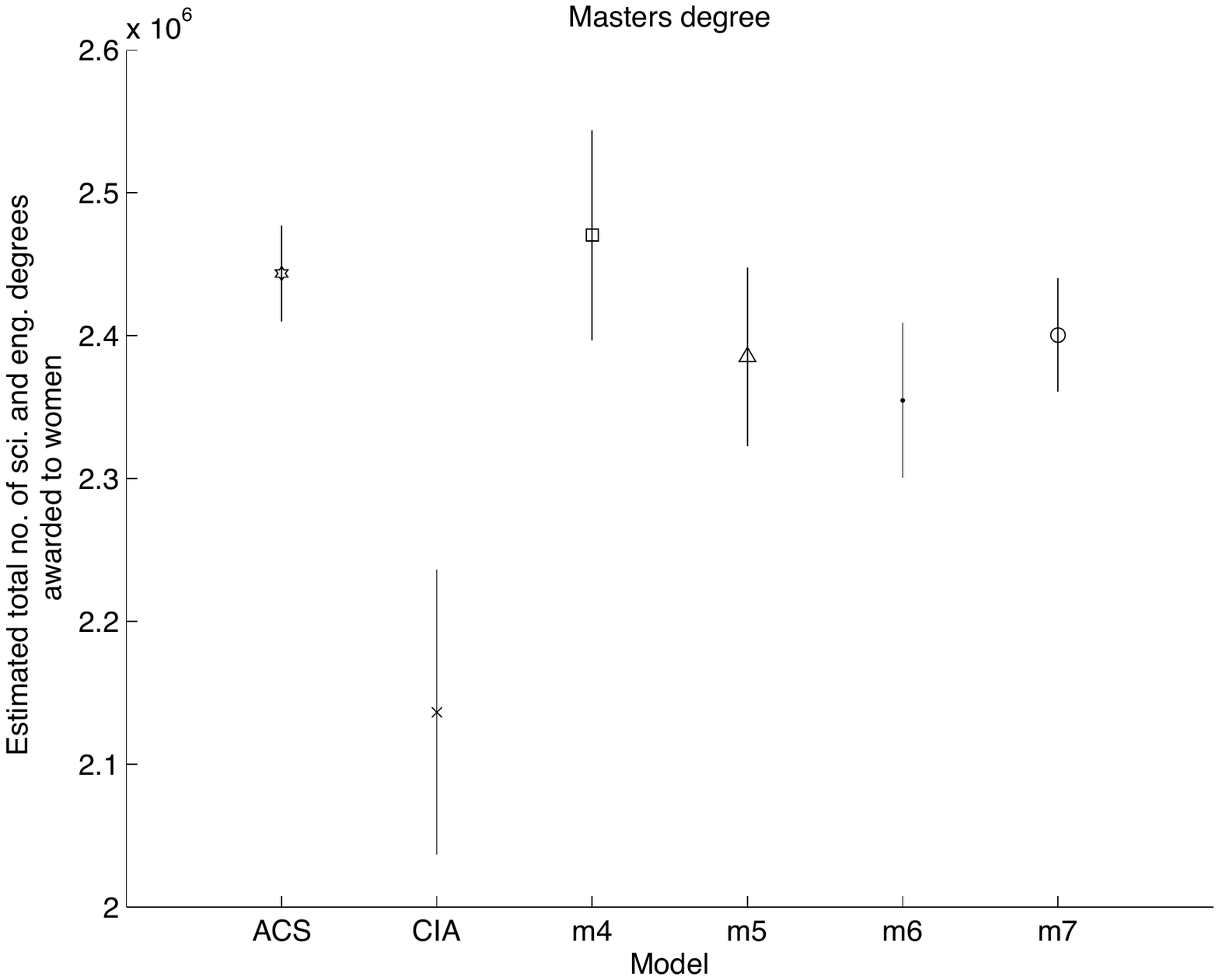}
\end{subfigure}
\begin{subfigure}{.48\textwidth}
\centering
\includegraphics[clip,trim=1cm 6cm 1cm 6cm,scale=.4]{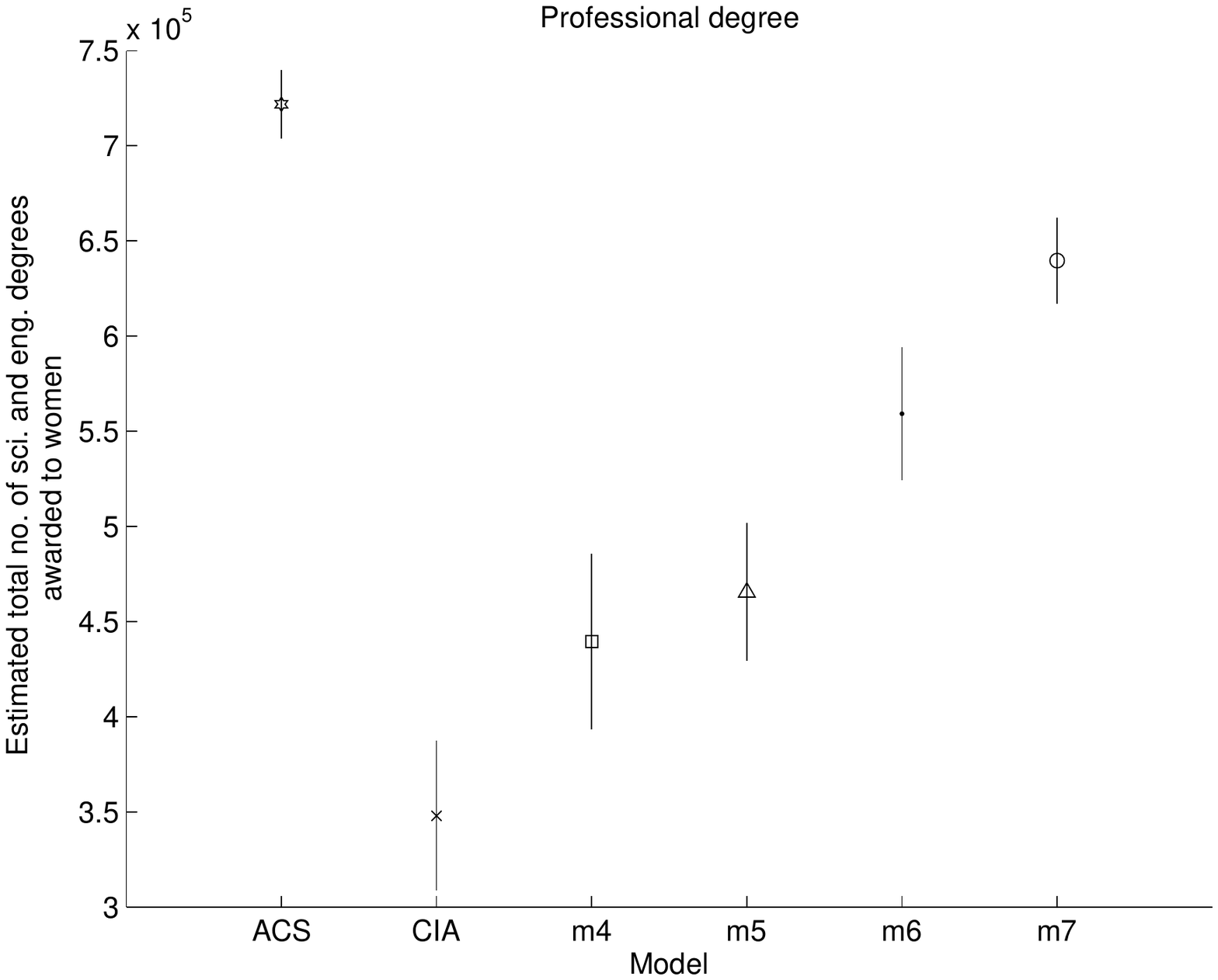}
\end{subfigure}
\begin{subfigure}{.48\textwidth}
\centering
\includegraphics[clip,trim=1cm 6cm 1cm 6cm,scale=.4]{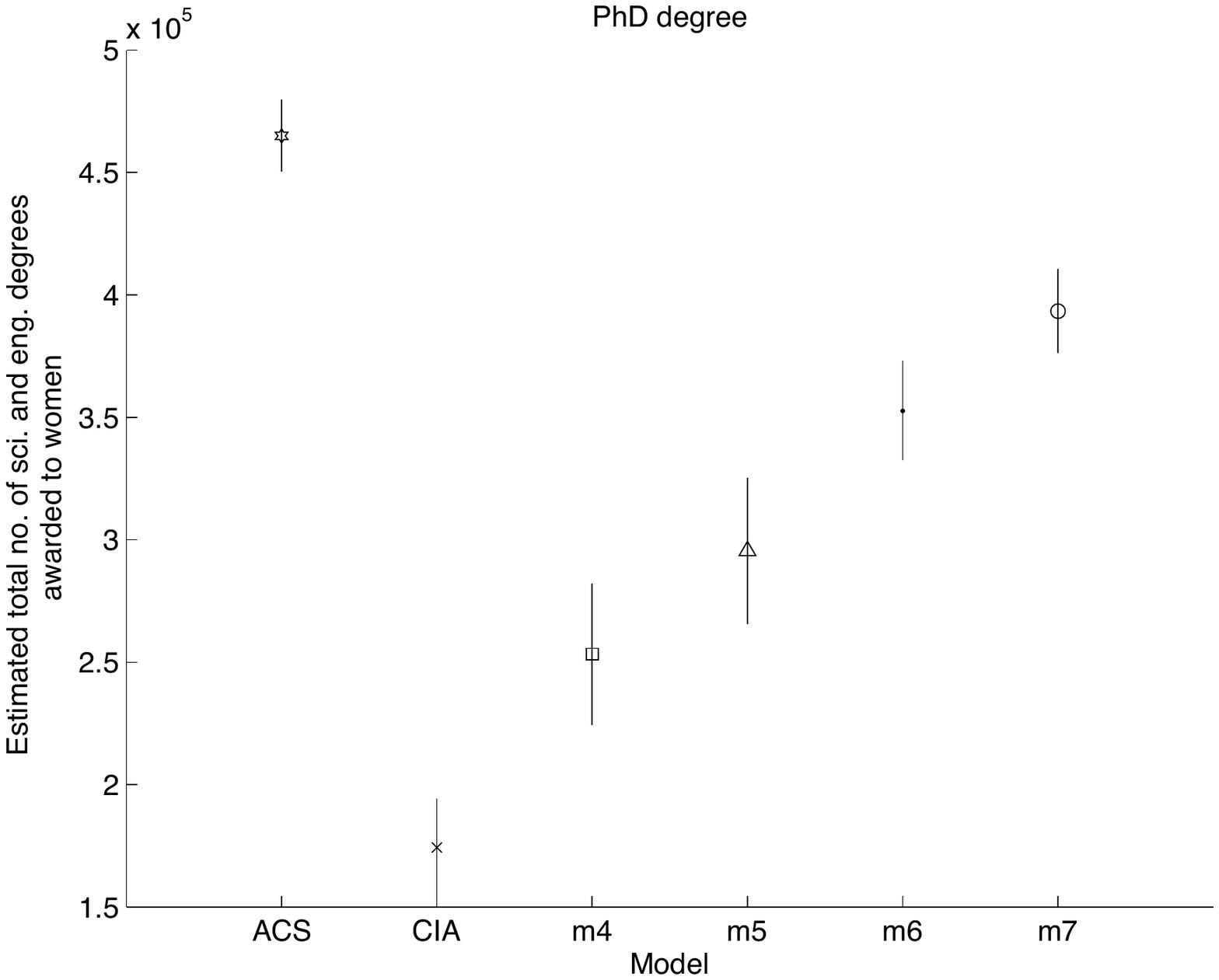}
\end{subfigure}
\caption[Total number of science and engineering degrees awarded to women.]{The estimated total number of science and engineering degrees awarded to women under each model. We plot the mean and 95$\%$ confidence intervals. Note the difference in scale for each degree category.}\label{fig:femalesci_total}
\end{figure}

Figure~\ref{fig:femalesci_total} displays the multiply-imputed, survey-weighted inferences for the total number of women with science and engineering degrees, computing using the ACS-specific indicator variable. We show results for Models 4 -- 7, the CIA model, and based on the ACS data without any adjustment for misreporting education.  The confidence intervals for Model 4 and Model 5 overlap substantially, suggesting not much practical difference in choosing among these models.  However, both are noticeably different from the other models, especially for the Ph.\ D.\ and professional degrees.  As the prior distributions on the error rates get stronger, the estimated counts increase towards the estimate using the ACS-reported education.  We note that using the ACS-reported education without adjustments results in substantially higher estimated totals at the professional and Ph.\ D.\ levels than any of the models that account for measurement error.  We also note that the CIA model yields considerably lower counts for all but bachelor's degrees.


\begin{figure}[t]
\begin{center}
\includegraphics[clip,trim=1cm 6cm 1cm 6cm, scale=.6]{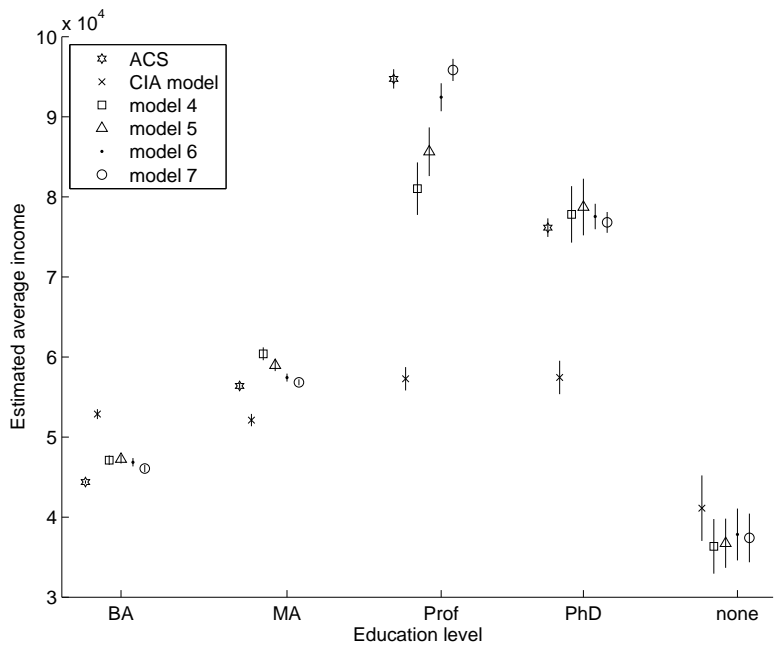}
\caption[Average income by education level.]{Multiple imputation point and 95\% confidence interval estimates for the average income within each education level. The ACS estimate is the survey-weighted estimate based on the reported education level in the 2010 ACS.}\label{fig:avgincome}
\end{center}
\end{figure}

Figure~\ref{fig:avgincome} displays inferences for the average income for different degrees. For most degrees, the point estimates for Models 4 -- 7 are reasonably close, with Models 4 and  5 again giving similar results.  The estimated average income for professionals differs noticeably across models, with Model 4 and Model 5 suggesting lower averages than the unadjusted ACS estimates, or than Models 6 and  7.   We note that the CIA model estimates are clearly implausible.  As an independent check on these estimates, we considered the estimated average earnings in the 2010 Current Population Survey.  They are \$83,720 for professional, \$80,600 for Ph.\ D.\ degree, \$66,144 for master's degree, and \$53,976 for bachelor's degree (http://www.collegequest.com/bls-research-education-pays-2010.aspx).  These line up more closely with the estimates from Model 5 than any other model, especially for the professional degree category, where the estimates most differ.

\begin{figure}[t]
\begin{subfigure}{.48\textwidth}
\centering
\includegraphics[clip,trim=1cm 6cm 1cm 6cm,scale=.4]{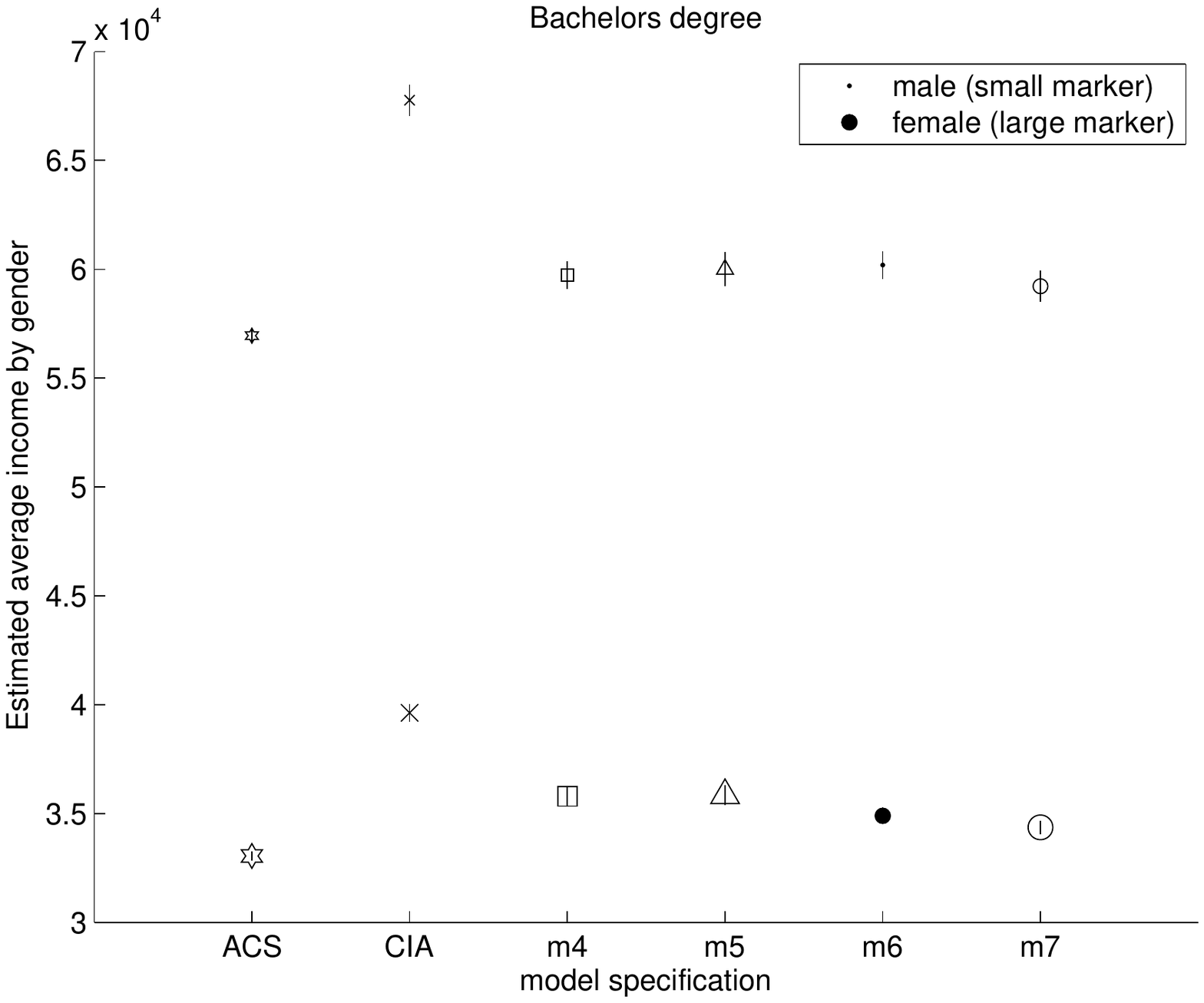}
\end{subfigure}
\begin{subfigure}{.48\textwidth}
\centering
\includegraphics[clip,trim=1cm 6cm 1cm 6cm,scale=.4]{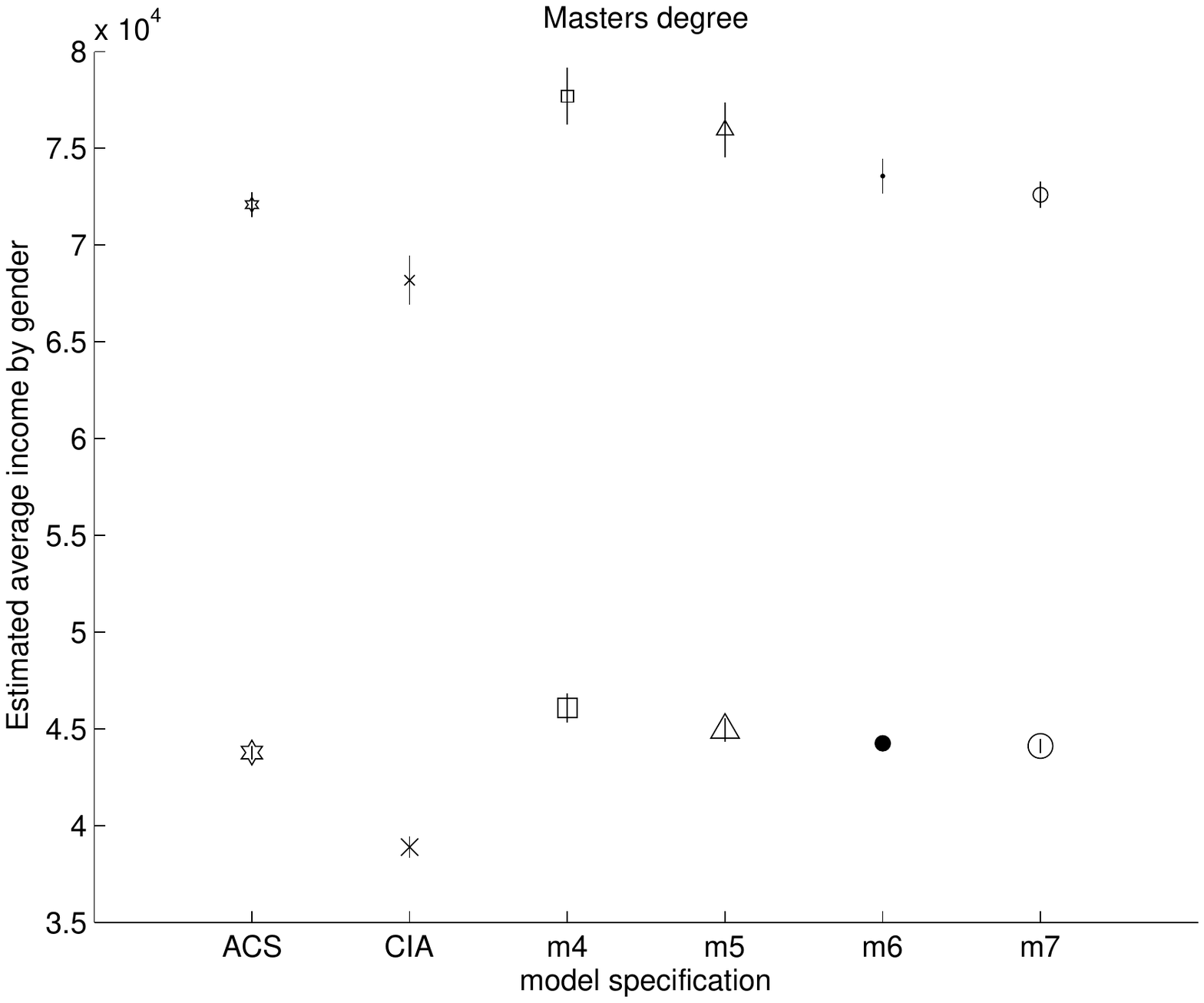}
\end{subfigure}
\begin{subfigure}{.48\textwidth}
\centering
\includegraphics[clip,trim=1cm 6cm 1cm 6cm,scale=.4]{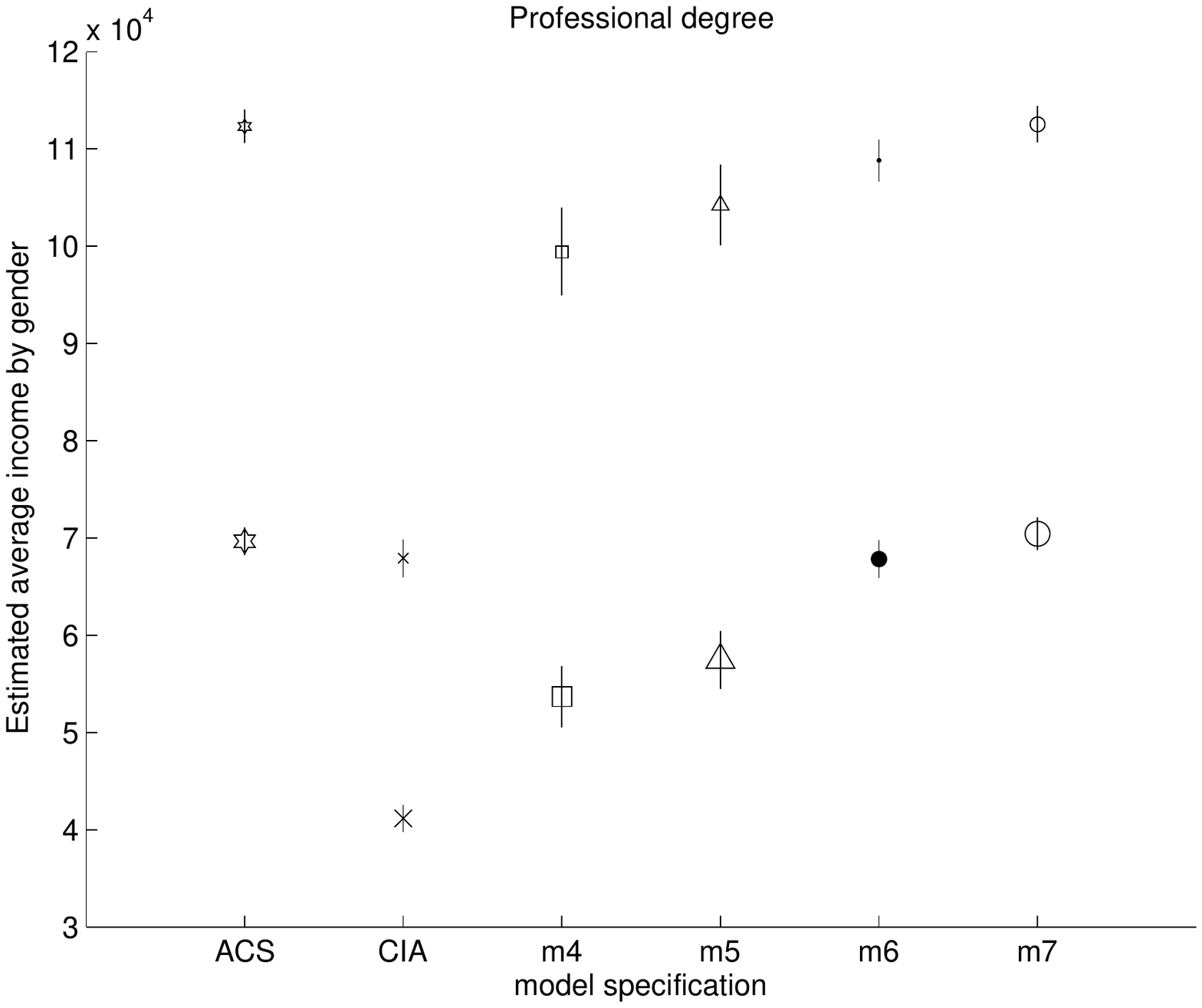}
\end{subfigure}
\begin{subfigure}{.48\textwidth}
\centering
\includegraphics[clip,trim=1cm 6cm 1cm 6cm,scale=.4]{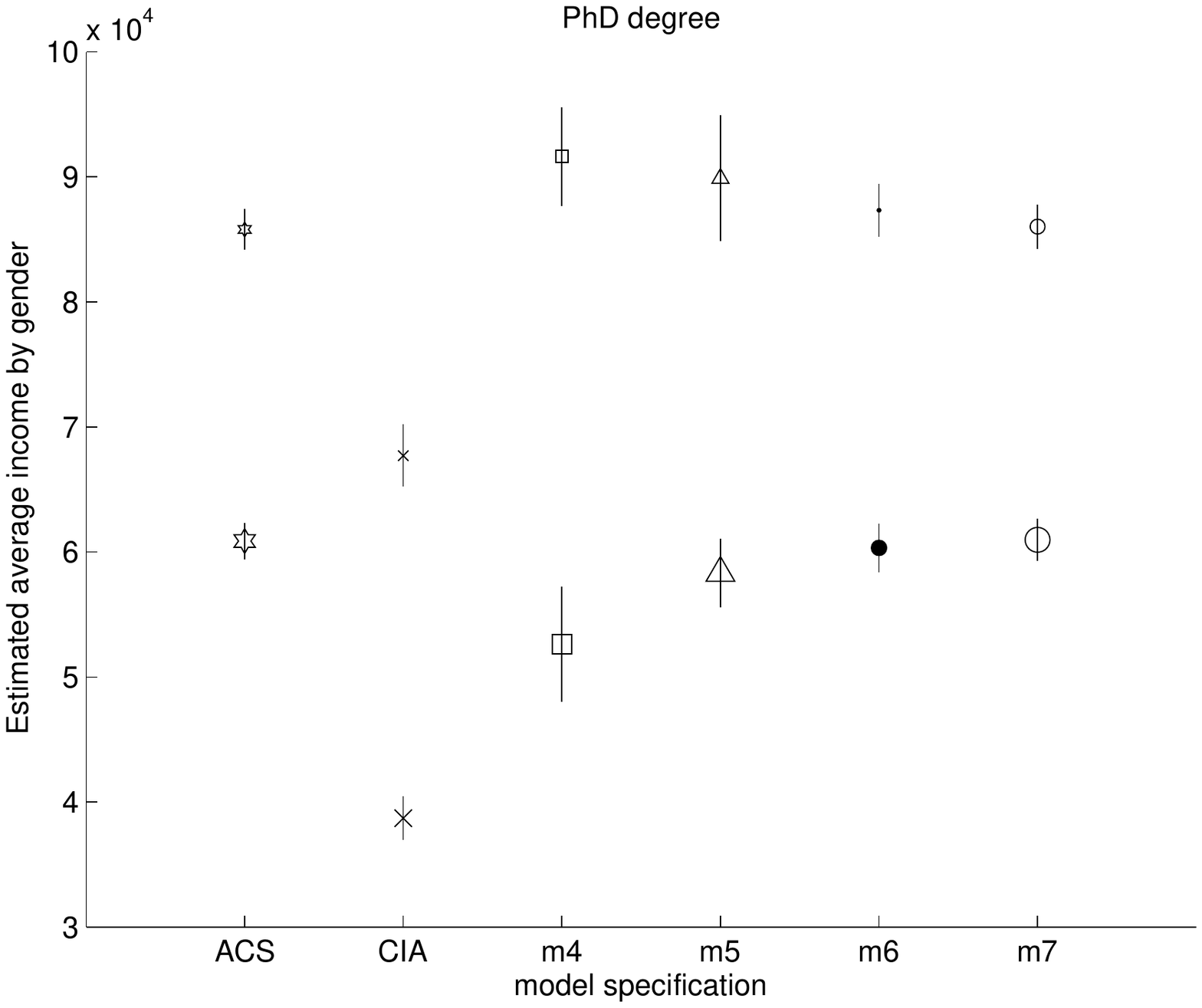}
\end{subfigure}
\caption[Difference between average male income and average female income.]{Multiple imputation point and 95\% confidence interval estimates for the average income for men and women within each education level. The ACS estimate is the survey-weighted estimate based on the reported education level in the 2010 ACS.}\label{fig:incomegap}
\end{figure}

Figure~\ref{fig:incomegap} displays inferences for the average income for men and women.  All models support the conclusion that men make more than women; apparently, misreporting in education does not account for that gap, at least for the models considered here.  We note that Model 4 suggests potentially larger income gaps between male and female Ph.\ D.\ recipients than the other models.

\section{Concluding Remarks}\label{sec:discussion}

The framework presented in this article offers analysts tools for using the information in a high quality, separate 
data source to adjust for measurement errors in the database of interest.  Key to the framework is to replace conditional independence assumptions typically used in data fusion with carefully considered measurement error models.  This avoids sacrificing information and facilitates analysis of the sensitivity of conclusions to alternative measurement error specifications.  Analysts 
can use diagnostic tests to rule out some measurement error models, and perform sensibility tests on others to identify 
reasonable candidates.

Besides survey sampling contexts like the one considered here involving the ACS and NSCG, the framework offers potential approaches for dealing with possible measurement errors in organic (big) data.  This is increasingly important, as 
data stewards and analysts consider replacing or supplementing high quality but expensive surveys with inexpensive and 
large-sample organic data.  Often, scant attention is paid to the potential impact of measurement
errors on inferences from those data.  The framework could be used with high quality, validated surveys as the gold standard data, 
allowing for adjustments to the error-prone organic data.

\section*{Supplementary Materials}
All supplemental files listed below are contained in a single .zip file (supplementary.zip) and can be obtained via a single download.

\begin{description}

\item[Supplementary Results:] Supplementary details and additional results for paper. (supp-material-final.pdf)

\item[ACS data:] 2010 ACS data used in the paper. (ACSdata\_2010standardized.csv.zip)

\item[Matlab code:] Matlab files containing main code MAINCODE\_edu\_2010app\_report1993.m and helper functions design.m and dirsamp.m, as well as parameter files mu.mat and tauSPD.mat. (code.zip)

\item[Prior Distributions:] Csv files are provided for priors used in Model 5 and read in by main Matlab code, referred to as femalereportprior1993.csv, malereportprior1993.csv, betareportprior.csv. (priors.zip)

\end{description}

\singlespacing
\bibliographystyle{asa}
\bibliography{References}

\begin{thebibliography}{33}
\newcommand{\enquote}[1]{``#1''}
\expandafter\ifx\csname natexlab\endcsname\relax\def\natexlab#1{#1}\fi

\bibitem[{Abayomi et~al.(2008)Abayomi, Gelman, and Levy}]{abayomi}
Abayomi, K., Gelman, A., and Levy, M. (2008), \enquote{Diagnostics for
  multivariate imputations,} \textit{Journal of the Royal Statistical Society:
  Series C (Applied Statistics)}, 57, 273--291.

\bibitem[{Black et~al.(2006)Black, Haviland, Sanders, and Taylor}]{sanders2006}
Black, D., Haviland, A., Sanders, S., and Taylor, L. (2006), \enquote{Why do
  minority men earn less? A study of wage differentials among the highly
  educated,} \textit{The Review of Economics and Statistics}, 88, 300--313.

\bibitem[{Black et~al.(2003)Black, Sanders, and Taylor}]{sanders2003}
Black, D., Sanders, S., and Taylor, L. (2003), \enquote{Measurement of higher
  education in the census and Current Population Survey,} \textit{Journal of
  the American Statistical Association}, 98, 545--554.

\bibitem[{Black et~al.(2008)Black, Haviland, Sanders, and Taylor}]{sanders2008}
Black, D.~A., Haviland, A.~M., Sanders, S.~G., and Taylor, L.~J. (2008),
  \enquote{Gender wage disparities among the highly educated,} \textit{Journal
  of Human Resources}, 43, 630--659.

\bibitem[{Carrig et~al.(2015)Carrig, Manrique-Vallier, Ranby, Reiter, and
  Hoyle}]{Carrig:2015}
Carrig, M., Manrique-Vallier, D., Ranby, K., Reiter, J.~P., and Hoyle, R.
  (2015), \enquote{A multiple imputation-based method for the retrospective
  harmonization of data sets,} \textit{Multivariate Behavioral Research}, 50,
  383--397.

\bibitem[{Curran and Hussong(2009)}]{curran2009}
Curran, P.~J. and Hussong, A.~M. (2009), \enquote{Integrative data analysis:
  The simultaneous analysis of multiple data sets,} \textit{Psychological
  Methods}, 14, 81--100.

\bibitem[{D'Orazio et~al.(2006)D'Orazio, Di~Zio, and Scanu}]{dorazio}
D'Orazio, M., Di~Zio, M., and Scanu, M. (2006), \textit{Statistical Matching:
  Theory and Practice}, Hoboken, NJ: Wiley.

\bibitem[{Dunson and Xing(2009)}]{dunson:xing:2009}
Dunson, D.~B. and Xing, C. (2009), \enquote{Nonparametric Bayes modeling of
  multivariate categorical data,} \textit{Journal of the American Statistical
  Association}, 104, 1042--1051.

\bibitem[{Fesco et~al.(2012)Fesco, Frase, and Kannankutty}]{ncses12201}
Fesco, R.~S., Frase, M.~J., and Kannankutty, N. (2012), \enquote{Using the
  American Community Survey as the sampling frame for the National Survey of
  College Graduates,} Working Paper NCSES 12-201, National Science Foundation,
  National Center for Science and Engineering Statistics, Arlington, VA.

\bibitem[{Finamore(2013)}]{nscgdoc}
Finamore, J. (2013), \textit{National Survey of College Graduates: About The
  Survey}, National Center for Science and Engineering Statistics.

\bibitem[{Fosdick et~al.(2016)Fosdick, DeYoreo, and Reiter}]{fosdick2015}
Fosdick, B.~K., DeYoreo, M., and Reiter, J.~P. (2016), \enquote{Categorical
  data fusion using auxiliary information,} \textit{Annals of Applied
  Statistics}, To appear.

\bibitem[{He et~al.(2014)He, Landrum, and Zaslavksy}]{he2014}
He, Y., Landrum, M.~B., and Zaslavksy, A.~M. (2014), \enquote{Combining
  information from two data sources with misreporting and incompleteness to
  assess hospice-use among cancer patients: a multiple imputation appraoch,}
  \textit{Statistics in Medicine}, 33, 3710--3724.

\bibitem[{Hirano et~al.(2001)Hirano, Imbens, Ridder, and
  Rubin}]{hirano2001combining}
Hirano, K., Imbens, G., Ridder, G., and Rubin, D. (2001), \enquote{{Combining
  panel data sets with attrition and refreshment samples},}
  \textit{Econometrica}, 69, 1645--1659.

\bibitem[{Kim et~al.(2015)Kim, Cox, Karr, Reiter, and Wang}]{kim2015}
Kim, H.~J., Cox, L.~H., Karr, A.~F., Reiter, J.~P., and Wang, Q. (2015),
  \enquote{Simultaneous edit-imputation for continuous microdata,}
  \textit{Journal of the American Statistical Association}, 110, 987 -- 999.

\bibitem[{Lohr(2010)}]{lohr}
Lohr, S.~L. (2010), \textit{Sampling: Design and Analysis}, Boston, MA:
  Brooks/Cole, 2nd ed.

\bibitem[{Manrique-Vallier and Reiter(2016)}]{daniel2015}
Manrique-Vallier, D. and Reiter, J.~P. (2016), \enquote{Bayesian simultaneous
  edit and imputation for multivariate categorical data,} \textit{Journal of
  the American Statistical Association}, To appear.

\bibitem[{Moriarity and Scheuren(2001)}]{moriarity}
Moriarity, C. and Scheuren, F. (2001), \enquote{Statistical matching: A
  paradigm for assessing the uncertainty in the procedure,} \textit{Journal of
  Official Statistics}, 17, 407 -- 422.

\bibitem[{{National Science Foundation}(1993)}]{icpsrdata}
{National Science Foundation} (1993), \enquote{National Survey of College
  Graduates, 1993,} http://doi.org/10.3886/ICPSR06880.v1, iCPSR06880-v1. Ann
  Arbor, MI: Inter-university Consortium for Political and Social Research
  [distributor], 2014-10-02.

\bibitem[{Pepe(1992)}]{pepe1992}
Pepe, M.~S. (1992), \enquote{Inference using surrogate outcome data and a
  validation sample,} \textit{Biometrika}, 79, 355 -- 365.

\bibitem[{Raghunathan(2006)}]{raghu2006}
Raghunathan, T.~E. (2006), \enquote{Combining information from multiple surveys
  for assessing health disparities,} \textit{Allgemeines Statistisches Archiv},
  90, 515--526.

\bibitem[{Rassler(2002)}]{rassler}
Rassler, S. (2002), \textit{Statistical Matching}, New York: Springer.

\bibitem[{Reiter(2008)}]{Reiter:Biometrika}
Reiter, J. (2008), \enquote{Multiple imputation when records used for
  imputation are not used or disseminated for analysis,} \textit{Biometrika},
  95, 933--946.

\bibitem[{Reiter(2012)}]{reiter:datafusion}
Reiter, J.~P. (2012), \enquote{Bayesian finite population imputation for data
  fusion,} \textit{Statistica Sinica}, 22, 795 -- 811.

\bibitem[{Rubin(1986)}]{rubinstatmatch}
Rubin, D.~B. (1986), \enquote{Statistical matching using file concatenation
  with adjusted weights and multiple imputations,} \textit{Journal of Business
  $\&$ Economic Statistics}, 4, 87--94.

\bibitem[{Rubin(1987)}]{rubin:1987}
--- (1987), \textit{Multiple Imputation for Nonresponse in Surveys}, New York:
  John Wiley $\&$ Sons.

\bibitem[{Schenker and Raghunathan(2007)}]{schenker2007}
Schenker, N. and Raghunathan, T.~E. (2007), \enquote{Combining information from
  multiple surveys to enhance estimation of measures of health,}
  \textit{Statistics in Medicine}, 26, 1802--1811.

\bibitem[{Schenker et~al.(2010)Schenker, Raghunathan, and
  Bondarenko}]{schenker2010}
Schenker, N., Raghunathan, T.~E., and Bondarenko, I. (2010), \enquote{Improving
  on analyses of self-reported data in a large-scale health survey by using
  information from an examination-based survey,} \textit{Statistics in
  Medicine}, 29, 533--545.

\bibitem[{Schifeling et~al.(2015)Schifeling, Cheng, Reiter, and
  Hillygus}]{refresh}
Schifeling, T.~A., Cheng, C., Reiter, J.~P., and Hillygus, D.~S. (2015),
  \enquote{Accounting for nonignorable unit nonresponse and attrition in panel
  studies with refreshment samples,} \textit{Journal of Survey Statistics and
  Methodology}, 3, 265 -- 295.

\bibitem[{Si and Reiter(2013)}]{si:reiter:jebs}
Si, Y. and Reiter, J. (2013), \enquote{Nonparametric Bayesian multiple
  imputation for incomplete categorical variables in large-scale assessment
  surveys,} \textit{Journal of Educational and Behavioral Statistics}, 38,
  499--521.

\bibitem[{Si et~al.(2015)Si, Reiter, and
  Hillygus}]{si:reiter:politicalanalysis}
Si, Y., Reiter, J.~P., and Hillygus, D.~S. (2015), \enquote{Semi-parametric
  selection models for potentially non-ignorable attrition in panel studies
  with Refreshment Samples,} \textit{Political Analysis}, 23, 92--112.

\bibitem[{Siddique et~al.(2015)Siddique, Reiter, Brincks, Gibbons, Crespi, and
  Brown}]{siddique}
Siddique, J., Reiter, J.~P., Brincks, A., Gibbons, R.~D., Crespi, C.~M., and
  Brown, C.~H. (2015), \enquote{Multiple imputation for harmonizing
  longitudinal non-commensurate measures in individual participant data
  meta-analysis,} \textit{Statistics in Medicine}, 34, 3399--3414.

\bibitem[{Tarmast(2001)}]{lognorm}
Tarmast, G. (2001), \enquote{Multivariate log-normal distribution,} in
  \textit{International Statistical Institute: Seoul 53rd Session}.

\bibitem[{Yucel and Zaslavsky(2005)}]{yucel2005}
Yucel, R.~M. and Zaslavsky, A.~M. (2005), \enquote{Imputation of binary
  treatment variables with measurement error in administrative data,}
  \textit{Journal of the American Statistical Association}, 100, 1123--1132.

\end{thebibliography}

\end{document}